\documentclass[aps, prl, reprint, 10pt,superscriptaddress]{revtex4-1}
\usepackage{graphicx}

\begin{document}

\title {Temperature and Field Dependence of Magnetic Domains in La$_{1.2}$Sr$_{1.8}$Mn$_2$O$_7$}
\author{Benjamin Bryant}
\affiliation{London Centre for Nanotechnology and Department of Physics and Astronomy, University College London, London WC1E 6BT, UK}
\affiliation{present address: Delft University of Technology, Kavli Institute of Nanoscience, Department of Quantum Nanoscience, Lorentzweg 1, 2628 CJ Delft, The Netherlands}
\author{Y. Moritomo}
\affiliation{Graduate School of Pure and Applied Science, University of Tsukuba, Tsukuba, Ibaraki 305-8571, Japan}
\author{Y. Tokura}
\affiliation{Multiferroic Project, ERATO, Japan Science and Technology Agency (JST), Wako, 351-0198, Japan}
\affiliation{Cross-Correlated Materials Research Group (CMRG), RIKEN, Advanced Science Institute, Wako, 351-0198, Japan}
\affiliation{Department of Applied Physics, University of Tokyo, Bunkyo-ku, Tokyo 113-8656, Japan}
\author{G. Aeppli}
\affiliation{London Centre for Nanotechnology and Department of Physics and Astronomy, University College London, London WC1E 6BT, UK}
\date{\today}

\begin{abstract}

Colossal magnetoresistance and field-induced ferromagnetism are well documented in manganite compounds. Since domain wall resistance contributes to magnetoresistance, data on the temperature and magnetic field dependence of the ferromagnetic domain structure are required for a full understanding of the magnetoresistive effect. Here we show, using cryogenic Magnetic Force Microscopy, domain structures for the layered manganite La$_{1.2}$Sr$_{1.8}$Mn$_2$O$_7$ as a function of temperature and magnetic field. Domain walls are suppressed close to the Curie temperature T$_C$, and appear either via the application of a c-axis magnetic field, or by decreasing the temperature further. At temperatures well below T$_C$, new domain walls, stable at zero field, can be formed by the application of a c-axis field. Magnetic structures are seen also at temperatures above T$_C$: these features are attributed to inclusions of additional Ruddleston-Popper manganite phases. Low-temperature domain walls are nucleated by these ferromagnetic inclusions.

\end{abstract}

\maketitle

Many manganite compounds exhibit negative colossal magnetoresistance (CMR), a very large reduction in electrical resistance upon application of a magnetic field \cite{CMR:review:Tokura}. Bilayer manganites exhibit colossal magnetoresistance in a similar way to the cubic compounds \cite{Tokura:Review}: the CMR effect appears to be enhanced by the bilayer structure \cite{Moritomo}. In all cases the largest magnetoresistance is found at temperatures close to the metal-insulator transition, which is attendant on the Curie transition. A simple phenomenological explanation for CMR is as an effect of spin disorder close to T$_C$. An applied magnetic field can suppress this disorder, enhancing the double-exchange hopping probability and hence the conductivity \cite{CMR:review:Tokura}. Effectively, the magnetic field polarizes the bands and therefore shifts the metal-insulator transition to a higher temperature. 

This simple explanation is obviously not sufficient: a complete model of colossal magnetoresistance in manganites must take into account effects such as phase separation \cite{Dagotto:PhaseSeparation:Review} where ferromagnetic metal regions are embedded in insulating matrices and vice versa, phenomena which have been studied in great detail for CMR manganites \cite{Murakami:nnano}. Phase separation can be both intrinsic and extrinsic, and given the complexity of the transition metal oxides, the latter must always be suspected. In particular, impurity phases with higher T$_C$ than the bulk will be critical for colossal magnetoresistance, as these will act as nucleation sites for the field-induced ferromagnetic phase. Ferromagnetic domain walls also contribute to magnetoresistance in the pseudocubic manganites \cite{Wu:domain, Mathur:domain}, particularly in ultra-thin films \cite{Li:domain}. Ferromagnetic domains in (La, Pr, Ca)MnO$_{3}$ have been imaged using Lorentz Microscopy (LTEM) \cite{Horibe2012, Schofield2012}, photoemission microscopy \cite{Burkhardt2012} and MFM \cite{Zhang2002}. Burkhardt et al. \cite{Burkhardt2012} were able to estimate the contribution of domain walls to magnetoresistance from field dependent LTEM images: this contribution is particularly important in the technologically-important low-field regime.  To understand magnetoresistance in layered manganites therefore, it is desirable that ferromagnetic domains be imaged both in the zero field low temperature state and in the field-induced ferromagnetic state. 

\begin{figure}[h]
\begin{centering}
\includegraphics{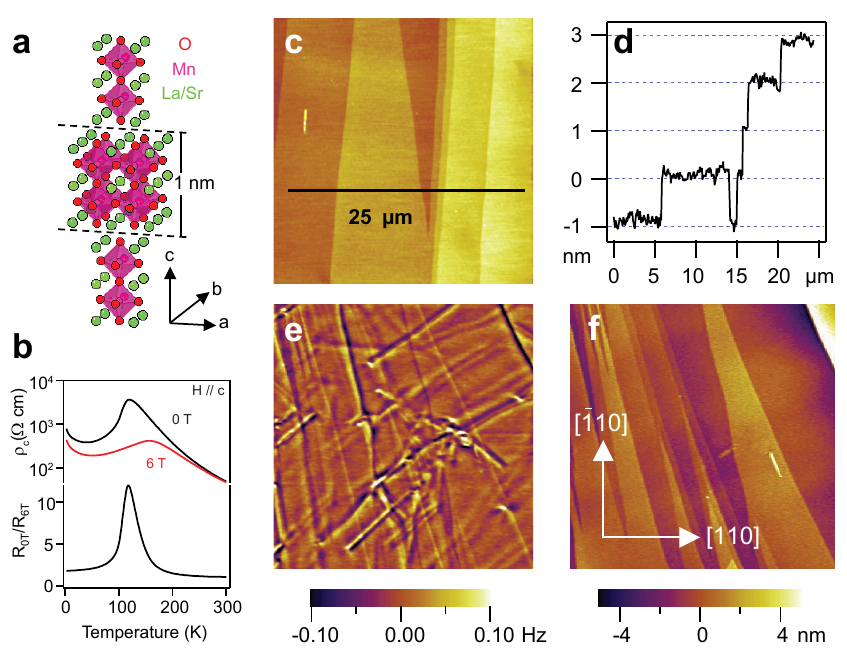}
\caption{\label{Figure1}(color online). (a) Crystal structure of La$_{1.2}$Sr$_{1.8}$Mn$_2$O$_7$ (b) C-axis resistivity as a function of temperature and c-axis field. Also shown is the magnetoresistance $R_{0T}/R_{6T}$. (c) Room temperature AFM topograph. The scale line shows the location of the cross section (d). (e) 25 x 25~$\mu$m MFM image collected at 4.7~K, showing domain walls. (f)~Corresponding AFM topographic image. }
\end{centering}
\end{figure}

To this end, we present here low-temperature Magnetic Force Microscopy (MFM) data for the ferromagnetic CMR bilayered manganite La$_{1.2}$Sr$_{1.8}$Mn$_{2}$O$_{7}$ \cite{Moritomo}. Bilayered manganites provide an opportunity to obtain good-quality surfaces, as these compounds may be readily cleaved to provide a clean, atomically flat surface \cite{Renner:LSMO, Bryant, Massee}. Previous spatially-resolved magnetic studies on bilayered manganites have included spin-polarized SEM on antiferromagnetic \cite{Konoto} and ferromagnetic \cite{Konoto:PRB} layered manganites, and MFM on the ferromagnet La$_{1.36}$Sr$_{1.64}$Mn$_{2}$O$_{7}$ (x=0.32) \cite{Huang}. 

Single crystal La$_{1.2}$Sr$_{1.8}$Mn$_{2}$O$_{7}$ samples were grown by an optical float zone method. Conductivity measurements confirmed the magnetoresistive effect: this peaks at 118 K, close to the metal-insulator transition (figure \ref{Figure1}b). Preliminary room-temperature AFM scans were carried out on La$_{1.2}$Sr$_{1.8}$Mn$_{2}$O$_{7}$ crystals, cleaved in air; figure \ref{Figure1}c shows a typical AFM topograph. The surface is largely clean and exhibits large terraces up to 10 $\mu$m across with a roughness of $<$ 0.1 nm. Terrace steps are always 1.0 $\pm$ 0.1 nm, or multiples thereof, corresponding to c/2 = 1.007 nm \cite{Moritomo}.

We used an Attocube low-temperature AFM for Magnetic Force Microscopy, in the temperature range 4.2~K to room temperature. The AFM was operated in Helium exchange gas, in frequency modulation mode. MFM images were obtained in units of frequency shift, $\Delta f \propto - \delta F_z / \delta z$, where $F_z$ is the z-component of the magnetic force between the tip and the sample stray field. Commercial MFM probes were used, with moment $\approx 0.3$x$10^{-13}$ e.m.u: the MFM lift height was 50 nm. A magnetic field of up to 8 T was applied, in the (vertical) c-axis direction normal to the sample surfaces. La$_{1.2}$Sr$_{1.8}$Mn$_{2}$O$_{7}$ single crystal samples were cleaved in air before being loaded into the low-temperature AFM. Bulk magnetization measurements were also carried out, using a Quantum Design SQUID magnetometer.

\begin{figure}
\begin{centering}
\includegraphics{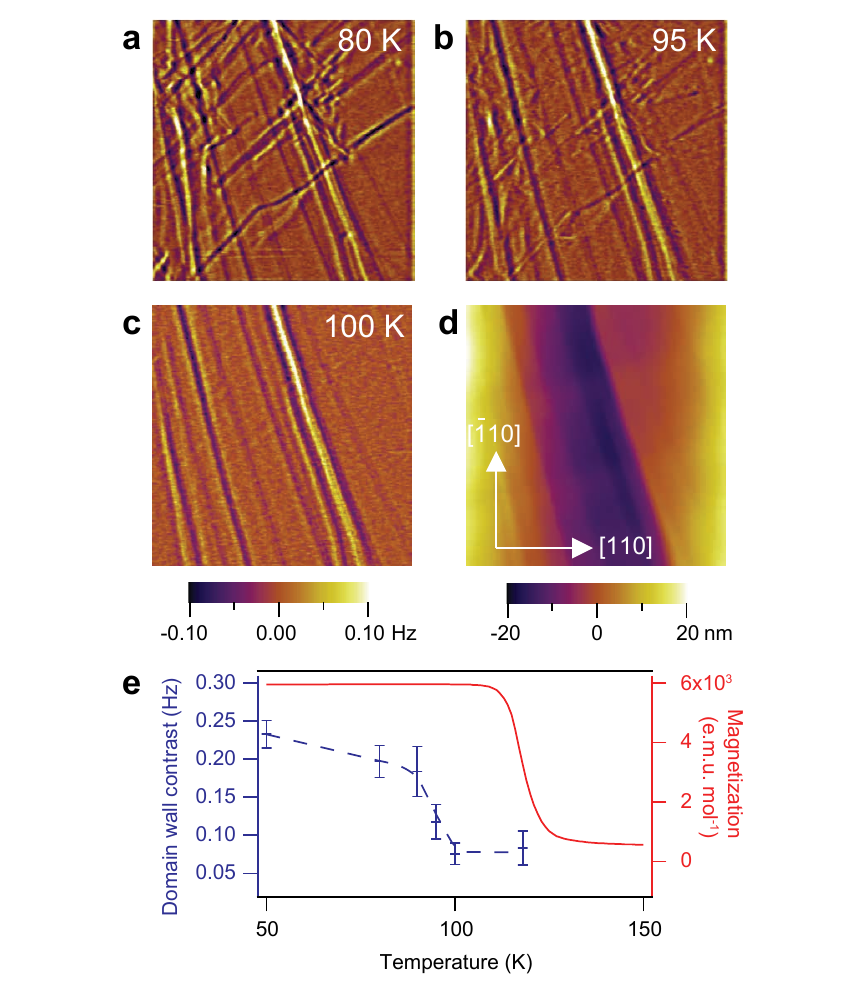}
\caption{\label{Figure2}(color online). 26 x 26 $\mu$m MFM images at 80~K (a), 95~K (b) and 100~K (c). Domain walls, visible at 80~K have disappeared at 100~K, leaving only topographic features. (d) AFM topographic image. (e) Comparison of bulk magnetization (H=100 Oe $\parallel$ ab) and surface domain wall contrast as observed by MFM, as functions of temperature: the dashed line is a guide to the eye. A steep drop in the visibility of domain walls is seen at 95~K, well below the bulk T$_C$ = 118~K.}
\end{centering}
\end{figure}

Figure \ref{Figure1}e is an MFM image of La$_{1.2}$Sr$_{1.8}$Mn$_{2}$O$_{7}$ collected at 4.7~K. Some crosstalk may be seen between the magnetic and topographic (figure \ref{Figure1}f) images, but the magnetic features are readily distinguished from terrace edges. The easy axis of magnetization is in the ab plane \cite{Hirota, Kubota}, so since the MFM tip is magnetized in the c-axis direction the magnetic contrast seen here is most likely due to Bloch-type domain walls. Linear domain walls are observed, with an average spacing of $\approx$ 5 $\mu$m: domain walls are observed to cross terrace edges, and are not aligned to the crystallographic axes. Some of the magnetic image features may represent two domain walls close together, i.e. a 2$\pi$ rotation of magnetic moment, for example those which are observed to terminate. Figure \ref{Figure2} shows a variable-temperature MFM study: the same area is imaged at 80~K, 95~K and 100~K. At 80~K the domain walls are clear, at 95~K they are still visible, but with reduced contrast, and by 100~K the domains are no longer visible. The remaining contrast at 100~K is due to topographic features (terrace edges). Figure \ref{Figure2}e shows the domain wall contrast, quantified as the peak to peak amplitude of the magnetic image, as a function of temperature in the range 50~K to 120~K. The effect of the topographic features on the measured amplitude has been eliminated by measuring sections parallel to the terrace edges. The bulk Curie temperature may be established as T$_C$ = 118~K from the onset of the low-field (100 Oe) magnetization, also shown in figure \ref{Figure2}e. The domain wall contrast sets in at a lower temperature, around 95~K. In previous MFM studies, domain wall contrast has been observed to increase with decreasing temperature below T$_C$ \cite{Soh:MMM, Ma:MMM, Lu:Science}, however these studies show a linear increase in contrast, rather than the sharp jump observed here. One possible explanation for the disappearance of magnetic contrast above 95~K is that above this temperature Bloch-type domain walls might transition to Ne\'el - capped domain walls \cite{Scheinfein:PRB} which would not be visible to MFM. This type of transition could occur due to the decrease of magnetic anisotropy close to T$_C$. Alternatively, a decrease in anisotropy might result in the stray field from the MFM tip overwriting the domain structure, so that it is no longer observed. In either case it is possible that above 95~K domain walls are still present, but not visible to MFM.

\begin{figure}
\begin{centering}
\includegraphics{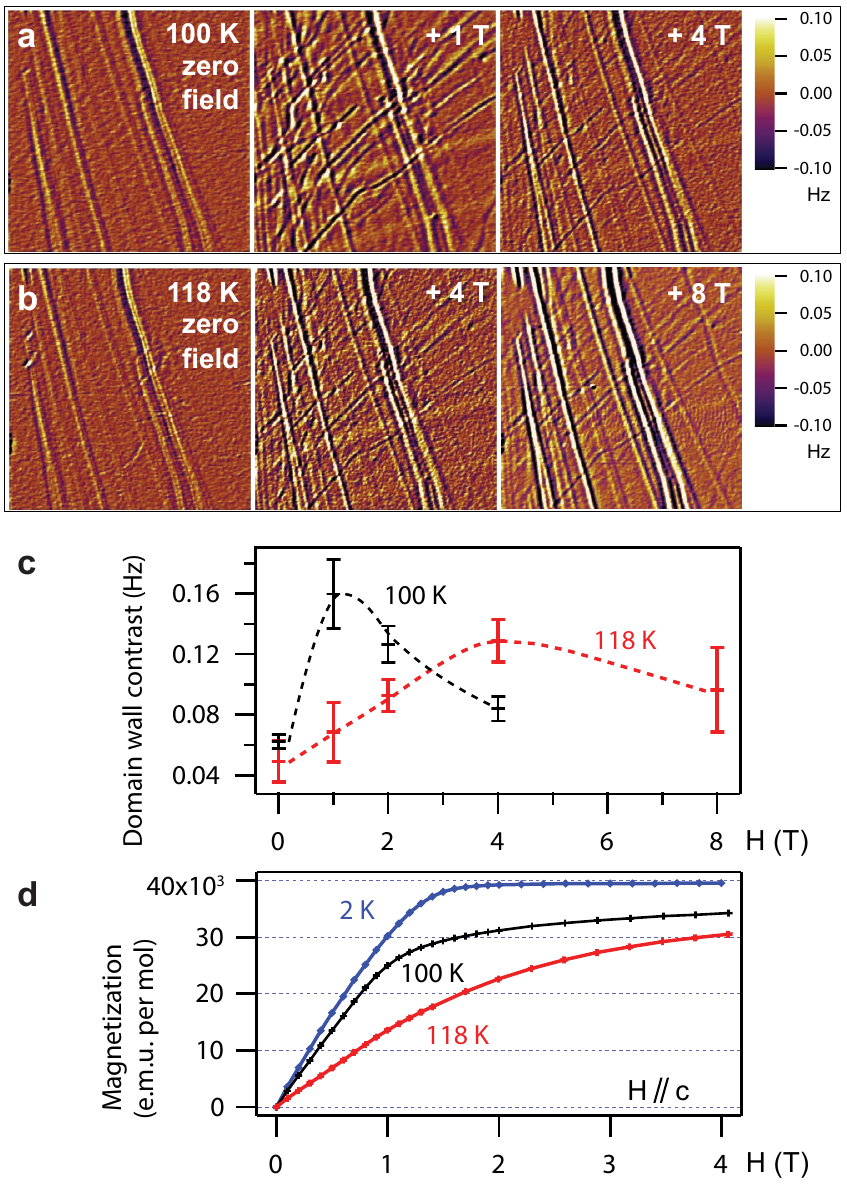}
\caption{\label{Figure3}(color online). (a) Field dependence of MFM imaging at 100~K:  the area is the same as for figure \ref{Figure2}c. (b) Field dependence of MFM imaging at 118~K: the area is the same as (a). All MFM images are 26 x 26 $\mu$m (c) Domain wall contrast as a function of field for 100~K and 118~K. Dashed lines are guides to the eye (d) Bulk magnetization vs. applied field (H $\parallel$ c) for 2K, 100~K and 118~K. Neither MFM data nor M vs. H have been corrected for the demagnetizing field, though both samples have a similar aspect ratio.}
\end{centering}
\end{figure}

As for other low-dimensional ferromagnets, La$_{1.2}$Sr$_{1.8}$Mn$_2$O$_7$ exhibits a large shift of the apparent T$_C$ to higher temperature upon application of a magnetic field (figure \ref{Figure1}b). We predict therefore that, in the temperature range 95~K $<$ T $<$ T$_C$, domain wall contrast will re-emerge with the application of field. By applying the field along the magnetically hard c-axis we may avoid completely magnetizing the sample, even at fields of several Tesla, enabling domain walls to be imaged at field. To this end magnetic field dependent MFM imaging was carried out at 118~K and 100~K. Figure \ref{Figure3}a shows the results of field-dependent MFM measurements at 100~K. The scan area is the same as in figure \ref{Figure2} and the field is applied along the hard c-axis. At zero field no domains are observed, while for an applied field of 1 T domains similar to those seen in the low-temperature state become visible. Comparison of figure \ref{Figure3}b to \ref{Figure2}a reveals that domain walls form in the same configuration under application of a field, as if the temperature is decreased. Thus an applied c-axis field mimics a decrease in temperature. At higher fields ($>$ 2 T) the domains become less clear, as the sample becomes fully magnetized along c. Figure \ref{Figure3}b shows the field dependent MFM images at 118~K. The result is similar to 100~K, but a much larger field is needed in order to make the domains visible, with peak domain contrast at 4 T. Figure \ref{Figure3}c summarizes the field dependence of the domain wall contrast, quantified as the peak to peak amplitude of the magnetic image, for 100~K and 118~K. Above a certain critical field the sample starts to become magnetized, and the domain contrast starts to decrease again: at both 100~K and 118~K the field-induced domain structure observed by MFM has maximum contrast when the sample magnetization has reached around 75 \% of the saturation value (figure \ref{Figure3}d). 

\begin{figure}
\begin{centering}
\includegraphics{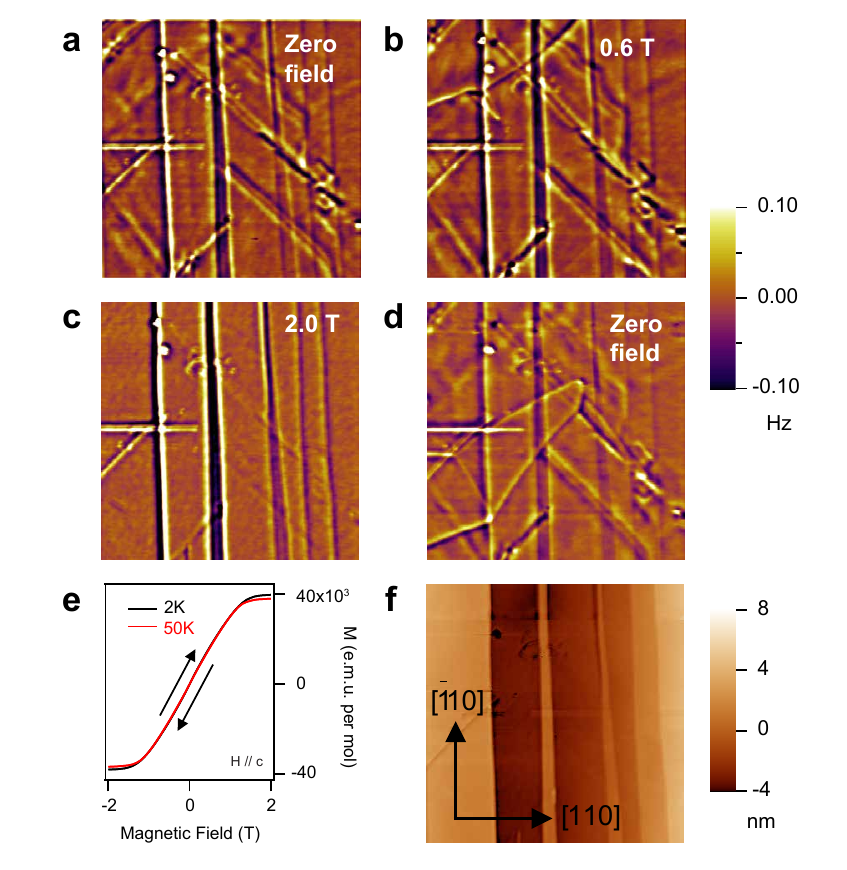}
\caption{\label{Figure4}(color online) Field dependence of MFM imaging at 20 K (a) Zero field MFM image, 15 x 15 $\mu$m. Some crosstalk with the topographic image is visible. Under 0.6 T applied field (b) a new domain wall is formed: this is wiped out by a field of 2.0 T (c). (d) Zero field MFM image, after a field of 8 T was applied. A new domain wall, stable at zero field, is observed. (e) M vs H for 2 K and 50 K, H $\parallel$ c, showing negligible hysteresis. (f) Topographic image of same area as a-d. All images 15 x 15 $\mu$m, all MFM images have the same color scale of $\pm$ 0.1 Hz.}
\end{centering}
\end{figure}

In the current experiment, because the field is applied perpendicular to the easy axis of magnetization, the energy to form Bloch walls is reduced by an applied field. This may be demonstrated by the formation of new domain walls under applied field, at temperatures well below T$_C$ (figure \ref{Figure4}). Figure \ref{Figure4}a shows domain structure at 20 K, at zero field. Upon the application of a 0.6 T field along the c-axis, a new domain wall is formed: this domain wall is observed to disappear at 2 T as the sample becomes magnetized. Zero-field imaging, after a field of 8 T was applied (figure \ref{Figure4}d) shows that a new domain wall has been formed.  Although the persistence of `new' domain walls at zero field implies some remanent magnetization, M vs. H curves (figure \ref{Figure4}e, see also \cite{Moritomo,Potter:PRB}) show negligible hysteresis, with coercivity $<$ 5 Oe. It is possible that remanent domains are purely a surface phenomenon, and make no substantial contribution to the bulk magnetization. 

\begin{figure}
\begin{centering}
\includegraphics[width=0.5\textwidth]{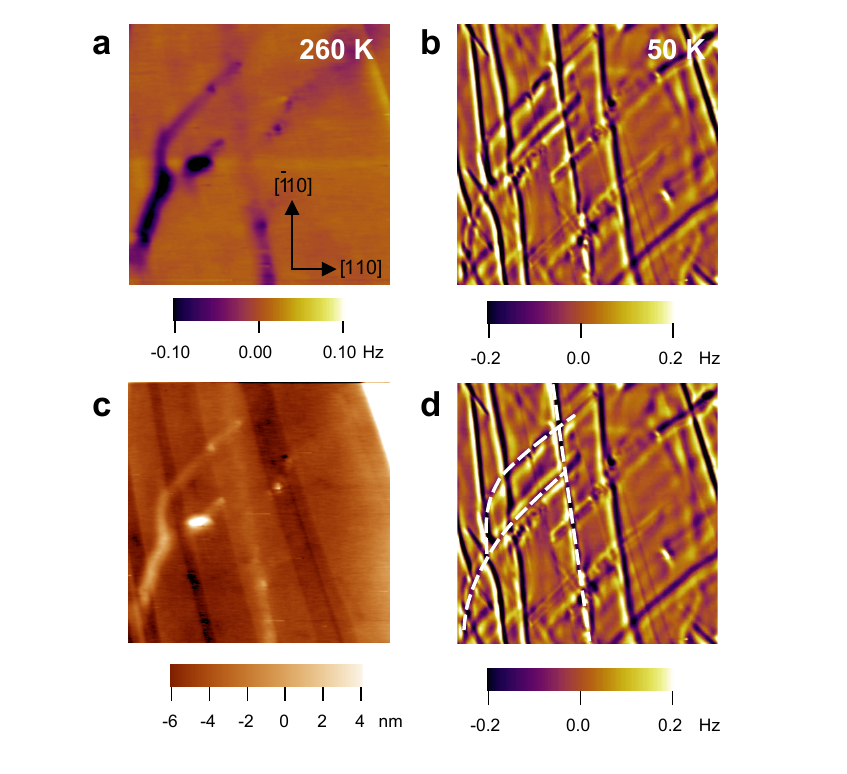}
\caption{\label{Figure5}(color online). Magnetic contrast above and below T$_C$ (a) 15 x 15 $\mu$m MFM image at 260~K (b) same area at 50~K. (c) Topographic image at 260~K of same area as (a) and (b), showing terraces edges, and also some crosstalk from the magnetic image. (d) Same as (b): dashed lines highlight magnetic features which persist above T$_C$.}
\end{centering}
\end{figure}

\begin{figure}
\begin{centering}
\includegraphics[width=0.5\textwidth]{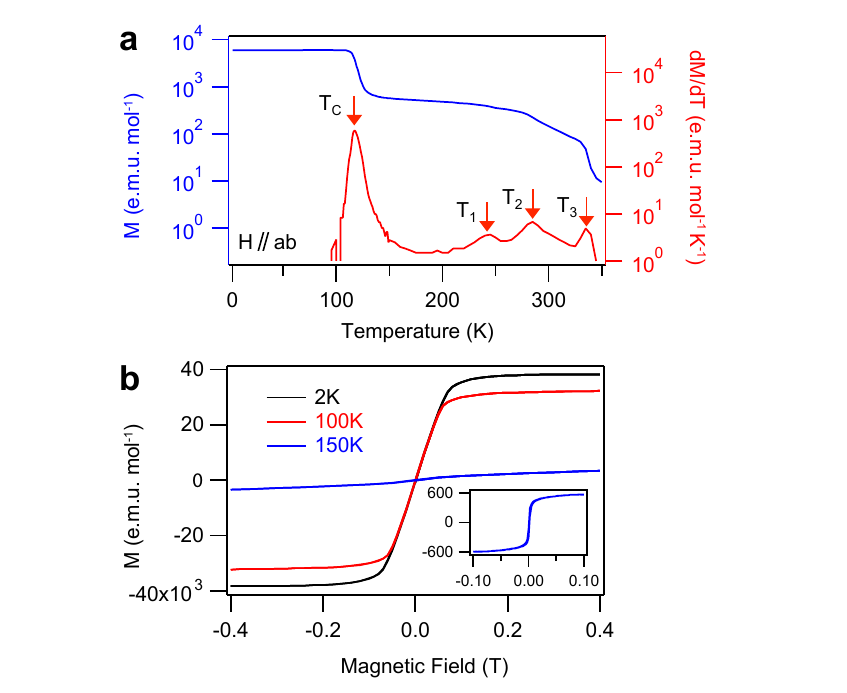}
\caption{\label{Figure6}(color online)(a) M vs T for the temperature range 2 K to 350 K. (H=100 Oe $\parallel$ ab). In addition to the bulk T$_C$ = 118 K additional magnetic transitions are observed at T$_1$ = 245 K, T$_2$ = 285 K and T$_3$ = 335 K. (b) M vs. H for 2 K, 100 K and 150 K. The inset shows M vs. H at 150 K with the paramagnetic background subtracted: a residual ferromagnetic component is observed.}
\end{centering}
\end{figure}

In a minority of locations on the La$_{1.2}$Sr$_{1.8}$Mn$_2$O$_7$ surface, magnetic image features are observed even well above T$_C$ = 118~K. Figure \ref{Figure5}a and b show MFM images of the same area at 260~K and 50~K: magnetic features are observed at 260~K as elongated structures 1-2 $\mu$m wide. Some crosstalk from the magnetic image can be seen in the topographic image (figure \ref{Figure5}c), however the features seen in \ref{Figure5}a can be positively identified as magnetic in origin since step edges seen in the topographic image are not seen in the MFM image. By comparison of the MFM images at 260~K and 50~K, it is clear that some magnetic features persist through T$_C$: figure \ref{Figure5}d highlights these features. Domain walls at 50~K are observed to form either as extensions of the magnetic features at 260~K or parallel to these features, suggesting that domains are nucleated by magnetic defects. The presence of an impurity phase with a higher Curie temperature in La$_{1.2}$Sr$_{1.8}$Mn$_2$O$_7$ may be inferred from bulk magnetization data. Figure \ref{Figure6}a shows M and dM/dT for an La$_{1.2}$Sr$_{1.8}$Mn$_2$O$_7$ sample from the same boule as MFM measurements. In addition to the bulk Curie transition at T$_C$ = 118~K, further higher temperature transitions are observed at T$_1$ = 245~K, T$_2$ = 285~K and T$_3$ = 335~K. In previous studies \cite{Potter:PRB, Allodi:PRB, Bader:JAP} such transitions at T $>$ T$_C$ have been attributed to intergrowths of n $>$ 2 variants of the Ruddleston-Popper series La$_{n-nx}$Sr$_{1+nx}$Mn$_n$O$_{3n+1}$. In general, for more three-dimensional compounds (higher~n), T$_C$ is higher: the cubic compound (n = $\infty$, La$_{0.6}$Sr$_{0.4}$MnO$_3$) has T$_C$ = 361~K \cite{Moritomo}. It is likely that the additional transitions at T$_1$, T$_2$ and T$_3$ represent different classes of inclusions with progressively higher n. The ratio of the saturation moment of the ferromagnetic component at T $>$ T$_C$ to the saturation moment at T $<$ T$_C$ \cite{Bader:JAP}, allows the volume fraction of inclusions to be estimated at 1.5 $\%$ (figure \ref{Figure6}b). The presence of n $>$ 2 impurity phases provides an explanation for the observation of magnetic features at T $>$ T$_C$: magnetic features in images such as figure \ref{Figure5}a indicate the location of such ferromagnetic inclusions. As the material is cooled below T$_C$ these inclusions act as nucleation points for the formation of domain walls.

Step heights of less than 1 nm, indicating the presence of n $\neq$ 2 phases at the surface, are not observed in AFM images of  La$_{1.2}$Sr$_{1.8}$Mn$_2$O$_7$. The magnetic features observed here for T $>$ T$_C$ therefore represent n $>$ 2 inclusions close to, but not at, the surface. Since cubic inclusions represent a small volume fraction of the material, and provide a less energetically favorable cleaving plane than the bulk bilayer structure \cite{Renner:BCMO,Loviat}, such phases are not expected to be observed directly at the cleaved surface. A cleave through an n $\neq$ 2 phase in La$_{2-2x}$Sr$_{1+2x}$Mn$_{2}$O$_{7}$ was observed by STM \cite{Massee}, but it was noted that this was unusual, being a single observation from a large number of cleaved surfaces.

In summary, we observe magnetic domain structures at low temperature in the ferromagnetic colossal magnetoresistive layered manganite La$_{1.2}$Sr$_{1.8}$Mn$_2$O$_7$. Upon increasing temperature, domain walls disappear at a temperature around 20~K below T$_C$, but may be observed to re-appear upon the application of a c-axis magnetic field. In addition, at temperatures well below T$_C$, the application of a uniform c-axis field causes new domain walls to be written to the material: these may be stable at zero field. We anticipate that these effects will have an impact on colossal magnetoresistance, due to the influence of domain wall resistance \cite{Wu:domain, Mathur:domain,Li:domain, Schofield2012}. Inclusions of n $>$ 2 Ruddleston-Popper phases in the layered material have been identified by magnetic imaging, since their transition temperatures are much higher than the bulk T$_C$. Upon cooling through T$_C$, domain walls are nucleated by these ferromagnetic inclusions.  Low-temperature MFM provides an ideal method to study magnetic phase inclusions and nucleation processes, both of which are crucial to a proper understanding of the phenomenon of colossal magnetoresistance.

\begin{acknowledgments}
The authors thank Kevin Heritage for assistance with the MFM setup, and Attocube Systems AG for technical support. 
\end{acknowledgments}

\bibliography{LSMObib}

\begin{thebibliography}{29}%
\makeatletter
\providecommand \@ifxundefined [1]{%
 \@ifx{#1\undefined}
}%
\providecommand \@ifnum [1]{%
 \ifnum #1\expandafter \@firstoftwo
 \else \expandafter \@secondoftwo
 \fi
}%
\providecommand \@ifx [1]{%
 \ifx #1\expandafter \@firstoftwo
 \else \expandafter \@secondoftwo
 \fi
}%
\providecommand \natexlab [1]{#1}%
\providecommand \enquote  [1]{``#1''}%
\providecommand \bibnamefont  [1]{#1}%
\providecommand \bibfnamefont [1]{#1}%
\providecommand \citenamefont [1]{#1}%
\providecommand \href@noop [0]{\@secondoftwo}%
\providecommand \href [0]{\begingroup \@sanitize@url \@href}%
\providecommand \@href[1]{\@@startlink{#1}\@@href}%
\providecommand \@@href[1]{\endgroup#1\@@endlink}%
\providecommand \@sanitize@url [0]{\catcode `\\12\catcode `\$12\catcode
  `\&12\catcode `\#12\catcode `\^12\catcode `\_12\catcode `\%12\relax}%
\providecommand \@@startlink[1]{}%
\providecommand \@@endlink[0]{}%
\providecommand \url  [0]{\begingroup\@sanitize@url \@url }%
\providecommand \@url [1]{\endgroup\@href {#1}{\urlprefix }}%
\providecommand \urlprefix  [0]{URL }%
\providecommand \Eprint [0]{\href }%
\providecommand \doibase [0]{http://dx.doi.org/}%
\providecommand \selectlanguage [0]{\@gobble}%
\providecommand \bibinfo  [0]{\@secondoftwo}%
\providecommand \bibfield  [0]{\@secondoftwo}%
\providecommand \translation [1]{[#1]}%
\providecommand \BibitemOpen [0]{}%
\providecommand \bibitemStop [0]{}%
\providecommand \bibitemNoStop [0]{.\EOS\space}%
\providecommand \EOS [0]{\spacefactor3000\relax}%
\providecommand \BibitemShut  [1]{\csname bibitem#1\endcsname}%
\let\auto@bib@innerbib\@empty
\bibitem [{\citenamefont {Tokura}(2006)}]{CMR:review:Tokura}%
  \BibitemOpen
  \bibfield  {author} {\bibinfo {author} {\bibfnamefont {Y.}~\bibnamefont
  {Tokura}},\ }\href {http://stacks.iop.org/0034-4885/69/i=3/a=R06} {\bibfield
  {journal} {\bibinfo  {journal} {Reports on Progress in Physics}\ }\textbf
  {\bibinfo {volume} {69}},\ \bibinfo {pages} {797} (\bibinfo {year}
  {2006})}\BibitemShut {NoStop}%
\bibitem [{\citenamefont {Kimura}\ and\ \citenamefont
  {Tokura}(2000)}]{Tokura:Review}%
  \BibitemOpen
  \bibfield  {author} {\bibinfo {author} {\bibfnamefont {T.}~\bibnamefont
  {Kimura}}\ and\ \bibinfo {author} {\bibfnamefont {Y.}~\bibnamefont
  {Tokura}},\ }\href@noop {} {\bibfield  {journal} {\bibinfo  {journal} {Annual
  Review of Materials Science}\ }\textbf {\bibinfo {volume} {30}},\ \bibinfo
  {pages} {451} (\bibinfo {year} {2000})}\BibitemShut {NoStop}%
\bibitem [{\citenamefont {Moritomo}\ \emph {et~al.}(1996)\citenamefont
  {Moritomo}, \citenamefont {Asamitsu}, \citenamefont {Kuwahara},\ and\
  \citenamefont {Tokura}}]{Moritomo}%
  \BibitemOpen
  \bibfield  {author} {\bibinfo {author} {\bibfnamefont {Y.}~\bibnamefont
  {Moritomo}}, \bibinfo {author} {\bibfnamefont {A.}~\bibnamefont {Asamitsu}},
  \bibinfo {author} {\bibfnamefont {H.}~\bibnamefont {Kuwahara}}, \ and\
  \bibinfo {author} {\bibfnamefont {Y.}~\bibnamefont {Tokura}},\ }\href@noop {}
  {\bibfield  {journal} {\bibinfo  {journal} {Nature}\ }\textbf {\bibinfo
  {volume} {380}},\ \bibinfo {pages} {141} (\bibinfo {year}
  {1996})}\BibitemShut {NoStop}%
\bibitem [{\citenamefont {Dagotto}\ \emph {et~al.}(2001)\citenamefont
  {Dagotto}, \citenamefont {Hotta},\ and\ \citenamefont
  {Moreo}}]{Dagotto:PhaseSeparation:Review}%
  \BibitemOpen
  \bibfield  {author} {\bibinfo {author} {\bibfnamefont {E.}~\bibnamefont
  {Dagotto}}, \bibinfo {author} {\bibfnamefont {T.}~\bibnamefont {Hotta}}, \
  and\ \bibinfo {author} {\bibfnamefont {A.}~\bibnamefont {Moreo}},\ }\href
  {\doibase 10.1016/S0370-1573(00)00121-6} {\bibfield  {journal} {\bibinfo
  {journal} {Physics Reports}\ }\textbf {\bibinfo {volume} {344}},\ \bibinfo
  {pages} {1} (\bibinfo {year} {2001})}\BibitemShut {NoStop}%
\bibitem [{\citenamefont {Murakami}\ \emph {et~al.}(2010)\citenamefont
  {Murakami}, \citenamefont {Kasai}, \citenamefont {Kim}, \citenamefont
  {Mamishin}, \citenamefont {Shindo}, \citenamefont {Mori},\ and\ \citenamefont
  {Tonomura}}]{Murakami:nnano}%
  \BibitemOpen
  \bibfield  {author} {\bibinfo {author} {\bibfnamefont {Y.}~\bibnamefont
  {Murakami}}, \bibinfo {author} {\bibfnamefont {H.}~\bibnamefont {Kasai}},
  \bibinfo {author} {\bibfnamefont {J.~J.}\ \bibnamefont {Kim}}, \bibinfo
  {author} {\bibfnamefont {S.}~\bibnamefont {Mamishin}}, \bibinfo {author}
  {\bibfnamefont {D.}~\bibnamefont {Shindo}}, \bibinfo {author} {\bibfnamefont
  {S.}~\bibnamefont {Mori}}, \ and\ \bibinfo {author} {\bibfnamefont
  {A.}~\bibnamefont {Tonomura}},\ }\href
  {http://dx.doi.org/10.1038/nnano.2009.342} {\bibfield  {journal} {\bibinfo
  {journal} {Nat Nano}\ }\textbf {\bibinfo {volume} {5}},\ \bibinfo {pages}
  {37} (\bibinfo {year} {2010})}\BibitemShut {NoStop}%
\bibitem [{\citenamefont {Wu}\ \emph {et~al.}(1999)\citenamefont {Wu},
  \citenamefont {Suzuki}, \citenamefont {Rudiger}, \citenamefont {Yu},
  \citenamefont {Kent}, \citenamefont {Nath},\ and\ \citenamefont
  {Eom}}]{Wu:domain}%
  \BibitemOpen
  \bibfield  {author} {\bibinfo {author} {\bibfnamefont {Y.}~\bibnamefont
  {Wu}}, \bibinfo {author} {\bibfnamefont {Y.}~\bibnamefont {Suzuki}}, \bibinfo
  {author} {\bibfnamefont {U.}~\bibnamefont {Rudiger}}, \bibinfo {author}
  {\bibfnamefont {J.}~\bibnamefont {Yu}}, \bibinfo {author} {\bibfnamefont
  {A.~D.}\ \bibnamefont {Kent}}, \bibinfo {author} {\bibfnamefont {T.~K.}\
  \bibnamefont {Nath}}, \ and\ \bibinfo {author} {\bibfnamefont {C.~B.}\
  \bibnamefont {Eom}},\ }\href {\doibase 10.1063/1.124995} {\bibfield
  {journal} {\bibinfo  {journal} {Applied Physics Letters}\ }\textbf {\bibinfo
  {volume} {75}},\ \bibinfo {pages} {2295} (\bibinfo {year}
  {1999})}\BibitemShut {NoStop}%
\bibitem [{\citenamefont {Mathur}\ \emph {et~al.}(1997)\citenamefont {Mathur},
  \citenamefont {Burnell}, \citenamefont {Isaac}, \citenamefont {Jackson},
  \citenamefont {Teo}, \citenamefont {MacManus-Driscoll}, \citenamefont
  {Cohen}, \citenamefont {Evetts},\ and\ \citenamefont
  {Blamire}}]{Mathur:domain}%
  \BibitemOpen
  \bibfield  {author} {\bibinfo {author} {\bibfnamefont {N.~D.}\ \bibnamefont
  {Mathur}}, \bibinfo {author} {\bibfnamefont {G.}~\bibnamefont {Burnell}},
  \bibinfo {author} {\bibfnamefont {S.~P.}\ \bibnamefont {Isaac}}, \bibinfo
  {author} {\bibfnamefont {T.~J.}\ \bibnamefont {Jackson}}, \bibinfo {author}
  {\bibfnamefont {B.~S.}\ \bibnamefont {Teo}}, \bibinfo {author} {\bibfnamefont
  {J.~L.}\ \bibnamefont {MacManus-Driscoll}}, \bibinfo {author} {\bibfnamefont
  {L.~F.}\ \bibnamefont {Cohen}}, \bibinfo {author} {\bibfnamefont {J.~E.}\
  \bibnamefont {Evetts}}, \ and\ \bibinfo {author} {\bibfnamefont {M.~G.}\
  \bibnamefont {Blamire}},\ }\href {http://dx.doi.org/10.1038/387266a0}
  {\bibfield  {journal} {\bibinfo  {journal} {Nature}\ }\textbf {\bibinfo
  {volume} {387}},\ \bibinfo {pages} {266} (\bibinfo {year}
  {1997})}\BibitemShut {NoStop}%
\bibitem [{\citenamefont {Li}\ \emph {et~al.}(2001)\citenamefont {Li},
  \citenamefont {Hu},\ and\ \citenamefont {Wang}}]{Li:domain}%
  \BibitemOpen
  \bibfield  {author} {\bibinfo {author} {\bibfnamefont {Q.}~\bibnamefont
  {Li}}, \bibinfo {author} {\bibfnamefont {Y.~F.}\ \bibnamefont {Hu}}, \ and\
  \bibinfo {author} {\bibfnamefont {H.~S.}\ \bibnamefont {Wang}},\ }\href
  {\doibase 10.1063/1.1362645} {\bibfield  {journal} {\bibinfo  {journal}
  {Journal of Applied Physics}\ }\textbf {\bibinfo {volume} {89}},\ \bibinfo
  {pages} {6952} (\bibinfo {year} {2001})}\BibitemShut {NoStop}%
\bibitem [{\citenamefont {Horibe}\ \emph {et~al.}(2012)\citenamefont {Horibe},
  \citenamefont {Mori}, \citenamefont {Asaka}, \citenamefont {Matsui},
  \citenamefont {Sharma}, \citenamefont {Koo}, \citenamefont {Guha},
  \citenamefont {Chen},\ and\ \citenamefont {Cheong}}]{Horibe2012}%
  \BibitemOpen
  \bibfield  {author} {\bibinfo {author} {\bibfnamefont {Y.}~\bibnamefont
  {Horibe}}, \bibinfo {author} {\bibfnamefont {S.}~\bibnamefont {Mori}},
  \bibinfo {author} {\bibfnamefont {T.}~\bibnamefont {Asaka}}, \bibinfo
  {author} {\bibfnamefont {Y.}~\bibnamefont {Matsui}}, \bibinfo {author}
  {\bibfnamefont {P.~a.}\ \bibnamefont {Sharma}}, \bibinfo {author}
  {\bibfnamefont {T.~Y.}\ \bibnamefont {Koo}}, \bibinfo {author} {\bibfnamefont
  {S.}~\bibnamefont {Guha}}, \bibinfo {author} {\bibfnamefont {C.~H.}\
  \bibnamefont {Chen}}, \ and\ \bibinfo {author} {\bibfnamefont {S.-W.}\
  \bibnamefont {Cheong}},\ }\href {\doibase 10.1209/0295-5075/100/67007}
  {\bibfield  {journal} {\bibinfo  {journal} {EPL (Europhysics Letters)}\
  }\textbf {\bibinfo {volume} {100}},\ \bibinfo {pages} {67007} (\bibinfo
  {year} {2012})}\BibitemShut {NoStop}%
\bibitem [{\citenamefont {Schofield}\ \emph {et~al.}(2012)\citenamefont
  {Schofield}, \citenamefont {He}, \citenamefont {Volkov},\ and\ \citenamefont
  {Zhu}}]{Schofield2012}%
  \BibitemOpen
  \bibfield  {author} {\bibinfo {author} {\bibfnamefont {M.~a.}\ \bibnamefont
  {Schofield}}, \bibinfo {author} {\bibfnamefont {J.}~\bibnamefont {He}},
  \bibinfo {author} {\bibfnamefont {V.~V.}\ \bibnamefont {Volkov}}, \ and\
  \bibinfo {author} {\bibfnamefont {Y.}~\bibnamefont {Zhu}},\ }\href {\doibase
  10.1063/1.4749396} {\bibfield  {journal} {\bibinfo  {journal} {Journal of
  Applied Physics}\ }\textbf {\bibinfo {volume} {112}},\ \bibinfo {pages}
  {053924} (\bibinfo {year} {2012})}\BibitemShut {NoStop}%
\bibitem [{\citenamefont {Burkhardt}\ \emph {et~al.}(2012)\citenamefont
  {Burkhardt}, \citenamefont {Hossain}, \citenamefont {Sarkar}, \citenamefont
  {Chuang}, \citenamefont {{Cruz Gonzalez}}, \citenamefont {Doran},
  \citenamefont {Scholl}, \citenamefont {Young}, \citenamefont {Tahir},
  \citenamefont {Choi}, \citenamefont {Cheong}, \citenamefont {D\"{u}rr},\ and\
  \citenamefont {St\"{o}hr}}]{Burkhardt2012}%
  \BibitemOpen
  \bibfield  {author} {\bibinfo {author} {\bibfnamefont {M.~H.}\ \bibnamefont
  {Burkhardt}}, \bibinfo {author} {\bibfnamefont {M.~a.}\ \bibnamefont
  {Hossain}}, \bibinfo {author} {\bibfnamefont {S.}~\bibnamefont {Sarkar}},
  \bibinfo {author} {\bibfnamefont {Y.-D.}\ \bibnamefont {Chuang}}, \bibinfo
  {author} {\bibfnamefont {A.~G.}\ \bibnamefont {{Cruz Gonzalez}}}, \bibinfo
  {author} {\bibfnamefont {A.}~\bibnamefont {Doran}}, \bibinfo {author}
  {\bibfnamefont {A.}~\bibnamefont {Scholl}}, \bibinfo {author} {\bibfnamefont
  {a.~T.}\ \bibnamefont {Young}}, \bibinfo {author} {\bibfnamefont
  {N.}~\bibnamefont {Tahir}}, \bibinfo {author} {\bibfnamefont {Y.~J.}\
  \bibnamefont {Choi}}, \bibinfo {author} {\bibfnamefont {S.-W.}\ \bibnamefont
  {Cheong}}, \bibinfo {author} {\bibfnamefont {H.~a.}\ \bibnamefont
  {D\"{u}rr}}, \ and\ \bibinfo {author} {\bibfnamefont {J.}~\bibnamefont
  {St\"{o}hr}},\ }\href
  {http://link.aps.org/doi/10.1103/PhysRevLett.108.237202} {\bibfield
  {journal} {\bibinfo  {journal} {Physical Review Letters}\ }\textbf {\bibinfo
  {volume} {108}},\ \bibinfo {pages} {237202} (\bibinfo {year}
  {2012})}\BibitemShut {NoStop}%
\bibitem [{\citenamefont {Zhang}\ \emph {et~al.}(2002)\citenamefont {Zhang},
  \citenamefont {Israel}, \citenamefont {Biswas}, \citenamefont {Greene},\ and\
  \citenamefont {de~Lozanne}}]{Zhang2002}%
  \BibitemOpen
  \bibfield  {author} {\bibinfo {author} {\bibfnamefont {L.}~\bibnamefont
  {Zhang}}, \bibinfo {author} {\bibfnamefont {C.}~\bibnamefont {Israel}},
  \bibinfo {author} {\bibfnamefont {A.}~\bibnamefont {Biswas}}, \bibinfo
  {author} {\bibfnamefont {R.~L.}\ \bibnamefont {Greene}}, \ and\ \bibinfo
  {author} {\bibfnamefont {A.}~\bibnamefont {de~Lozanne}},\ }\href {\doibase
  10.1126/science.1077346} {\bibfield  {journal} {\bibinfo  {journal} {Science
  (New York, N.Y.)}\ }\textbf {\bibinfo {volume} {298}},\ \bibinfo {pages}
  {805} (\bibinfo {year} {2002})}\BibitemShut {NoStop}%
\bibitem [{\citenamefont {Ronnow}\ \emph {et~al.}(2006)\citenamefont {Ronnow},
  \citenamefont {Renner}, \citenamefont {Aeppli}, \citenamefont {Kimura},\ and\
  \citenamefont {Tokura}}]{Renner:LSMO}%
  \BibitemOpen
  \bibfield  {author} {\bibinfo {author} {\bibfnamefont {H.}~\bibnamefont
  {Ronnow}}, \bibinfo {author} {\bibfnamefont {C.}~\bibnamefont {Renner}},
  \bibinfo {author} {\bibfnamefont {G.}~\bibnamefont {Aeppli}}, \bibinfo
  {author} {\bibfnamefont {T.}~\bibnamefont {Kimura}}, \ and\ \bibinfo {author}
  {\bibfnamefont {Y.}~\bibnamefont {Tokura}},\ }\href {\doibase
  10.1038/nature04650} {\bibfield  {journal} {\bibinfo  {journal} {Nature}\
  }\textbf {\bibinfo {volume} {440}},\ \bibinfo {pages} {1025} (\bibinfo {year}
  {2006})}\BibitemShut {NoStop}%
\bibitem [{\citenamefont {Bryant}\ \emph {et~al.}(2011)\citenamefont {Bryant},
  \citenamefont {Renner}, \citenamefont {Tokunaga}, \citenamefont {Tokura},\
  and\ \citenamefont {Aeppli}}]{Bryant}%
  \BibitemOpen
  \bibfield  {author} {\bibinfo {author} {\bibfnamefont {B.}~\bibnamefont
  {Bryant}}, \bibinfo {author} {\bibfnamefont {C.}~\bibnamefont {Renner}},
  \bibinfo {author} {\bibfnamefont {Y.}~\bibnamefont {Tokunaga}}, \bibinfo
  {author} {\bibfnamefont {Y.}~\bibnamefont {Tokura}}, \ and\ \bibinfo {author}
  {\bibfnamefont {G.}~\bibnamefont {Aeppli}},\ }\href {\doibase
  10.1038/ncomms1219} {\bibfield  {journal} {\bibinfo  {journal} {Nature
  Communications}\ }\textbf {\bibinfo {volume} {2}} (\bibinfo {year} {2011}),\
  10.1038/ncomms1219}\BibitemShut {NoStop}%
\bibitem [{\citenamefont {Massee}\ \emph {et~al.}(2011)\citenamefont {Massee},
  \citenamefont {de~Jong}, \citenamefont {Huang}, \citenamefont {Siu},
  \citenamefont {Santoso}, \citenamefont {Mans}, \citenamefont {Boothroyd},
  \citenamefont {Prabhakaran}, \citenamefont {Follath}, \citenamefont
  {Varykhalov}, \citenamefont {Patthey}, \citenamefont {Shi}, \citenamefont
  {Goedkoop},\ and\ \citenamefont {Golden}}]{Massee}%
  \BibitemOpen
  \bibfield  {author} {\bibinfo {author} {\bibfnamefont {F.}~\bibnamefont
  {Massee}}, \bibinfo {author} {\bibfnamefont {S.}~\bibnamefont {de~Jong}},
  \bibinfo {author} {\bibfnamefont {Y.}~\bibnamefont {Huang}}, \bibinfo
  {author} {\bibfnamefont {W.~K.}\ \bibnamefont {Siu}}, \bibinfo {author}
  {\bibfnamefont {I.}~\bibnamefont {Santoso}}, \bibinfo {author} {\bibfnamefont
  {A.}~\bibnamefont {Mans}}, \bibinfo {author} {\bibfnamefont {A.~T.}\
  \bibnamefont {Boothroyd}}, \bibinfo {author} {\bibfnamefont {D.}~\bibnamefont
  {Prabhakaran}}, \bibinfo {author} {\bibfnamefont {R.}~\bibnamefont
  {Follath}}, \bibinfo {author} {\bibfnamefont {A.}~\bibnamefont {Varykhalov}},
  \bibinfo {author} {\bibfnamefont {L.}~\bibnamefont {Patthey}}, \bibinfo
  {author} {\bibfnamefont {M.}~\bibnamefont {Shi}}, \bibinfo {author}
  {\bibfnamefont {J.~B.}\ \bibnamefont {Goedkoop}}, \ and\ \bibinfo {author}
  {\bibfnamefont {M.~S.}\ \bibnamefont {Golden}},\ }\href
  {http://dx.doi.org/10.1038/nphys2089} {\bibfield  {journal} {\bibinfo
  {journal} {Nat Phys}\ }\textbf {\bibinfo {volume} {7}},\ \bibinfo {pages}
  {978} (\bibinfo {year} {2011})}\BibitemShut {NoStop}%
\bibitem [{\citenamefont {Konoto}\ \emph {et~al.}(2004)\citenamefont {Konoto},
  \citenamefont {Kohashi}, \citenamefont {Koike}, \citenamefont {Arima},
  \citenamefont {Kaneko}, \citenamefont {Kimura},\ and\ \citenamefont
  {Tokura}}]{Konoto}%
  \BibitemOpen
  \bibfield  {author} {\bibinfo {author} {\bibfnamefont {M.}~\bibnamefont
  {Konoto}}, \bibinfo {author} {\bibfnamefont {T.}~\bibnamefont {Kohashi}},
  \bibinfo {author} {\bibfnamefont {K.}~\bibnamefont {Koike}}, \bibinfo
  {author} {\bibfnamefont {T.}~\bibnamefont {Arima}}, \bibinfo {author}
  {\bibfnamefont {Y.}~\bibnamefont {Kaneko}}, \bibinfo {author} {\bibfnamefont
  {T.}~\bibnamefont {Kimura}}, \ and\ \bibinfo {author} {\bibfnamefont
  {Y.}~\bibnamefont {Tokura}},\ }\href@noop {} {\bibfield  {journal} {\bibinfo
  {journal} {Physical Review Letters}\ }\textbf {\bibinfo {volume} {93}}
  (\bibinfo {year} {2004})}\BibitemShut {NoStop}%
\bibitem [{\citenamefont {Konoto}\ \emph {et~al.}(2005)\citenamefont {Konoto},
  \citenamefont {Kohashi}, \citenamefont {Koike}, \citenamefont {Arima},
  \citenamefont {Kaneko}, \citenamefont {Kimura},\ and\ \citenamefont
  {Tokura}}]{Konoto:PRB}%
  \BibitemOpen
  \bibfield  {author} {\bibinfo {author} {\bibfnamefont {M.}~\bibnamefont
  {Konoto}}, \bibinfo {author} {\bibfnamefont {T.}~\bibnamefont {Kohashi}},
  \bibinfo {author} {\bibfnamefont {K.}~\bibnamefont {Koike}}, \bibinfo
  {author} {\bibfnamefont {T.}~\bibnamefont {Arima}}, \bibinfo {author}
  {\bibfnamefont {Y.}~\bibnamefont {Kaneko}}, \bibinfo {author} {\bibfnamefont
  {T.}~\bibnamefont {Kimura}}, \ and\ \bibinfo {author} {\bibfnamefont
  {Y.}~\bibnamefont {Tokura}},\ }\href@noop {} {\bibfield  {journal} {\bibinfo
  {journal} {Physical Review B}\ }\textbf {\bibinfo {volume} {71}} (\bibinfo
  {year} {2005})}\BibitemShut {NoStop}%
\bibitem [{\citenamefont {Huang}\ \emph {et~al.}(2008)\citenamefont {Huang},
  \citenamefont {Hyun}, \citenamefont {Chuang}, \citenamefont {Kim},
  \citenamefont {Goodenough}, \citenamefont {Zhou}, \citenamefont {Mitchell},\
  and\ \citenamefont {de~Lozanne}}]{Huang}%
  \BibitemOpen
  \bibfield  {author} {\bibinfo {author} {\bibfnamefont {J.}~\bibnamefont
  {Huang}}, \bibinfo {author} {\bibfnamefont {C.}~\bibnamefont {Hyun}},
  \bibinfo {author} {\bibfnamefont {T.-M.}\ \bibnamefont {Chuang}}, \bibinfo
  {author} {\bibfnamefont {J.}~\bibnamefont {Kim}}, \bibinfo {author}
  {\bibfnamefont {J.~B.}\ \bibnamefont {Goodenough}}, \bibinfo {author}
  {\bibfnamefont {J.~S.}\ \bibnamefont {Zhou}}, \bibinfo {author}
  {\bibfnamefont {J.~F.}\ \bibnamefont {Mitchell}}, \ and\ \bibinfo {author}
  {\bibfnamefont {A.}~\bibnamefont {de~Lozanne}},\ }\href {\doibase
  10.1103/PhysRevB.77.024405} {\bibfield  {journal} {\bibinfo  {journal} {Phys.
  Rev. B}\ }\textbf {\bibinfo {volume} {77}} (\bibinfo {year} {2008}),\
  10.1103/PhysRevB.77.024405}\BibitemShut {NoStop}%
\bibitem [{\citenamefont {Hirota}\ \emph {et~al.}(1998)\citenamefont {Hirota},
  \citenamefont {Moritomo}, \citenamefont {Fujioka}, \citenamefont {Kubota},
  \citenamefont {Yoshizawa},\ and\ \citenamefont {Endoh}}]{Hirota}%
  \BibitemOpen
  \bibfield  {author} {\bibinfo {author} {\bibfnamefont {K.}~\bibnamefont
  {Hirota}}, \bibinfo {author} {\bibfnamefont {Y.}~\bibnamefont {Moritomo}},
  \bibinfo {author} {\bibfnamefont {H.}~\bibnamefont {Fujioka}}, \bibinfo
  {author} {\bibfnamefont {M.}~\bibnamefont {Kubota}}, \bibinfo {author}
  {\bibfnamefont {H.}~\bibnamefont {Yoshizawa}}, \ and\ \bibinfo {author}
  {\bibfnamefont {Y.}~\bibnamefont {Endoh}},\ }\href {\doibase
  10.1143/JPSJ.67.3380} {\bibfield  {journal} {\bibinfo  {journal} {J. Phys.
  Soc. Jpn.}\ }\textbf {\bibinfo {volume} {67}},\ \bibinfo {pages} {3380}
  (\bibinfo {year} {1998})}\BibitemShut {NoStop}%
\bibitem [{\citenamefont {Kubota}\ \emph {et~al.}(1999)\citenamefont {Kubota},
  \citenamefont {Fujioka}, \citenamefont {Ohoyama}, \citenamefont {Hirota},
  \citenamefont {Moritomo}, \citenamefont {Yoshizawa},\ and\ \citenamefont
  {Endoh}}]{Kubota}%
  \BibitemOpen
  \bibfield  {author} {\bibinfo {author} {\bibfnamefont {M.}~\bibnamefont
  {Kubota}}, \bibinfo {author} {\bibfnamefont {H.}~\bibnamefont {Fujioka}},
  \bibinfo {author} {\bibfnamefont {K.}~\bibnamefont {Ohoyama}}, \bibinfo
  {author} {\bibfnamefont {K.}~\bibnamefont {Hirota}}, \bibinfo {author}
  {\bibfnamefont {Y.}~\bibnamefont {Moritomo}}, \bibinfo {author}
  {\bibfnamefont {H.}~\bibnamefont {Yoshizawa}}, \ and\ \bibinfo {author}
  {\bibfnamefont {Y.}~\bibnamefont {Endoh}},\ }\href@noop {} {\bibfield
  {journal} {\bibinfo  {journal} {Journal of Physics and Chemistry of Solids}\
  }\textbf {\bibinfo {volume} {60}},\ \bibinfo {pages} {1161} (\bibinfo {year}
  {1999})}\BibitemShut {NoStop}%
\bibitem [{\citenamefont {Soh}\ \emph {et~al.}(2001)\citenamefont {Soh},
  \citenamefont {Aeppli}, \citenamefont {Mathur},\ and\ \citenamefont
  {Blamire}}]{Soh:MMM}%
  \BibitemOpen
  \bibfield  {author} {\bibinfo {author} {\bibfnamefont {Y.}~\bibnamefont
  {Soh}}, \bibinfo {author} {\bibfnamefont {G.}~\bibnamefont {Aeppli}},
  \bibinfo {author} {\bibfnamefont {N.}~\bibnamefont {Mathur}}, \ and\ \bibinfo
  {author} {\bibfnamefont {M.}~\bibnamefont {Blamire}},\ }\href {\doibase
  10.1016/S0304-8853(00)01060-X} {\bibfield  {journal} {\bibinfo  {journal}
  {Journal of Magnetism and Magnetic Materials}\ }\textbf {\bibinfo {volume}
  {226}},\ \bibinfo {pages} {857} (\bibinfo {year} {2001})}\BibitemShut
  {NoStop}%
\bibitem [{\citenamefont {Ma}\ \emph {et~al.}(2002)\citenamefont {Ma},
  \citenamefont {Chueh}, \citenamefont {Kuang}, \citenamefont {Liou},\ and\
  \citenamefont {Yao}}]{Ma:MMM}%
  \BibitemOpen
  \bibfield  {author} {\bibinfo {author} {\bibfnamefont {Y.}~\bibnamefont
  {Ma}}, \bibinfo {author} {\bibfnamefont {C.}~\bibnamefont {Chueh}}, \bibinfo
  {author} {\bibfnamefont {W.}~\bibnamefont {Kuang}}, \bibinfo {author}
  {\bibfnamefont {Y.}~\bibnamefont {Liou}}, \ and\ \bibinfo {author}
  {\bibfnamefont {Y.}~\bibnamefont {Yao}},\ }\href {\doibase
  10.1016/S0304-8853(01)00606-0} {\bibfield  {journal} {\bibinfo  {journal}
  {Journal of Magnetism and Magnetic Materials}\ }\textbf {\bibinfo {volume}
  {239}},\ \bibinfo {pages} {371} (\bibinfo {year} {2002})}\BibitemShut
  {NoStop}%
\bibitem [{\citenamefont {Lu}\ \emph {et~al.}(1997)\citenamefont {Lu},
  \citenamefont {Chen},\ and\ \citenamefont {de~Lozanne}}]{Lu:Science}%
  \BibitemOpen
  \bibfield  {author} {\bibinfo {author} {\bibfnamefont {Q.}~\bibnamefont
  {Lu}}, \bibinfo {author} {\bibfnamefont {C.}~\bibnamefont {Chen}}, \ and\
  \bibinfo {author} {\bibfnamefont {A.}~\bibnamefont {de~Lozanne}},\ }\href
  {\doibase 10.1126/science.276.5321.2006} {\bibfield  {journal} {\bibinfo
  {journal} {Science}\ }\textbf {\bibinfo {volume} {276}},\ \bibinfo {pages}
  {2006} (\bibinfo {year} {1997})}\BibitemShut {NoStop}%
\bibitem [{\citenamefont {Scheinfein}\ \emph {et~al.}(1991)\citenamefont
  {Scheinfein}, \citenamefont {Unguris}, \citenamefont {Blue}, \citenamefont
  {Coakley}, \citenamefont {Pierce}, \citenamefont {Celotta},\ and\
  \citenamefont {Ryan}}]{Scheinfein:PRB}%
  \BibitemOpen
  \bibfield  {author} {\bibinfo {author} {\bibfnamefont {M.~R.}\ \bibnamefont
  {Scheinfein}}, \bibinfo {author} {\bibfnamefont {J.}~\bibnamefont {Unguris}},
  \bibinfo {author} {\bibfnamefont {J.~L.}\ \bibnamefont {Blue}}, \bibinfo
  {author} {\bibfnamefont {K.~J.}\ \bibnamefont {Coakley}}, \bibinfo {author}
  {\bibfnamefont {D.~T.}\ \bibnamefont {Pierce}}, \bibinfo {author}
  {\bibfnamefont {R.~J.}\ \bibnamefont {Celotta}}, \ and\ \bibinfo {author}
  {\bibfnamefont {P.~J.}\ \bibnamefont {Ryan}},\ }\href {\doibase
  10.1103/PhysRevB.43.3395} {\bibfield  {journal} {\bibinfo  {journal}
  {Physical Review B}\ }\textbf {\bibinfo {volume} {43}},\ \bibinfo {pages}
  {3395} (\bibinfo {year} {1991})}\BibitemShut {NoStop}%
\bibitem [{\citenamefont {Potter}\ \emph {et~al.}(1998)\citenamefont {Potter},
  \citenamefont {Swiatek}, \citenamefont {Bader}, \citenamefont {Argyriou},
  \citenamefont {Mitchell}, \citenamefont {Miller}, \citenamefont {Hinks},\
  and\ \citenamefont {Jorgensen}}]{Potter:PRB}%
  \BibitemOpen
  \bibfield  {author} {\bibinfo {author} {\bibfnamefont {C.}~\bibnamefont
  {Potter}}, \bibinfo {author} {\bibfnamefont {M.}~\bibnamefont {Swiatek}},
  \bibinfo {author} {\bibfnamefont {S.}~\bibnamefont {Bader}}, \bibinfo
  {author} {\bibfnamefont {D.}~\bibnamefont {Argyriou}}, \bibinfo {author}
  {\bibfnamefont {J.}~\bibnamefont {Mitchell}}, \bibinfo {author}
  {\bibfnamefont {D.}~\bibnamefont {Miller}}, \bibinfo {author} {\bibfnamefont
  {D.}~\bibnamefont {Hinks}}, \ and\ \bibinfo {author} {\bibfnamefont
  {J.}~\bibnamefont {Jorgensen}},\ }\href {\doibase 10.1103/PhysRevB.57.72}
  {\bibfield  {journal} {\bibinfo  {journal} {Physical Review B}\ }\textbf
  {\bibinfo {volume} {57}},\ \bibinfo {pages} {72} (\bibinfo {year}
  {1998})}\BibitemShut {NoStop}%
\bibitem [{\citenamefont {Allodi}\ \emph {et~al.}(2008)\citenamefont {Allodi},
  \citenamefont {Bimbi}, \citenamefont {De~Renzi}, \citenamefont {Baumann},
  \citenamefont {Apostu}, \citenamefont {Suryanarayanan},\ and\ \citenamefont
  {Revcolevschi}}]{Allodi:PRB}%
  \BibitemOpen
  \bibfield  {author} {\bibinfo {author} {\bibfnamefont {G.}~\bibnamefont
  {Allodi}}, \bibinfo {author} {\bibfnamefont {M.}~\bibnamefont {Bimbi}},
  \bibinfo {author} {\bibfnamefont {R.}~\bibnamefont {De~Renzi}}, \bibinfo
  {author} {\bibfnamefont {C.}~\bibnamefont {Baumann}}, \bibinfo {author}
  {\bibfnamefont {M.}~\bibnamefont {Apostu}}, \bibinfo {author} {\bibfnamefont
  {R.}~\bibnamefont {Suryanarayanan}}, \ and\ \bibinfo {author} {\bibfnamefont
  {A.}~\bibnamefont {Revcolevschi}},\ }\href {\doibase
  10.1103/PhysRevB.78.064420} {\bibfield  {journal} {\bibinfo  {journal}
  {Physical Review B}\ }\textbf {\bibinfo {volume} {78}},\ \bibinfo {pages}
  {064420} (\bibinfo {year} {2008})}\BibitemShut {NoStop}%
\bibitem [{\citenamefont {Bader}\ \emph {et~al.}(1998)\citenamefont {Bader},
  \citenamefont {Osgood}, \citenamefont {Miller}, \citenamefont {Mitchell},\
  and\ \citenamefont {Jiang}}]{Bader:JAP}%
  \BibitemOpen
  \bibfield  {author} {\bibinfo {author} {\bibfnamefont {S.}~\bibnamefont
  {Bader}}, \bibinfo {author} {\bibfnamefont {R.}~\bibnamefont {Osgood}},
  \bibinfo {author} {\bibfnamefont {D.}~\bibnamefont {Miller}}, \bibinfo
  {author} {\bibfnamefont {J.}~\bibnamefont {Mitchell}}, \ and\ \bibinfo
  {author} {\bibfnamefont {J.}~\bibnamefont {Jiang}},\ }\href {\doibase
  10.1063/1.367906} {\bibfield  {journal} {\bibinfo  {journal} {Journal of
  Applied Physics}\ }\textbf {\bibinfo {volume} {83}},\ \bibinfo {pages} {6385}
  (\bibinfo {year} {1998})}\BibitemShut {NoStop}%
\bibitem [{\citenamefont {Renner}\ \emph {et~al.}(2002)\citenamefont {Renner},
  \citenamefont {Aeppli}, \citenamefont {Kim}, \citenamefont {Soh},\ and\
  \citenamefont {Cheong}}]{Renner:BCMO}%
  \BibitemOpen
  \bibfield  {author} {\bibinfo {author} {\bibfnamefont {C.}~\bibnamefont
  {Renner}}, \bibinfo {author} {\bibfnamefont {G.}~\bibnamefont {Aeppli}},
  \bibinfo {author} {\bibfnamefont {B.}~\bibnamefont {Kim}}, \bibinfo {author}
  {\bibfnamefont {Y.}~\bibnamefont {Soh}}, \ and\ \bibinfo {author}
  {\bibfnamefont {S.}~\bibnamefont {Cheong}},\ }\href@noop {} {\bibfield
  {journal} {\bibinfo  {journal} {Nature}\ }\textbf {\bibinfo {volume} {416}},\
  \bibinfo {pages} {518} (\bibinfo {year} {2002})}\BibitemShut {NoStop}%
\bibitem [{\citenamefont {Loviat}\ \emph {et~al.}(2007)\citenamefont {Loviat},
  \citenamefont {Ronnow}, \citenamefont {Renner}, \citenamefont {Aeppli},
  \citenamefont {Kimura},\ and\ \citenamefont {Tokura}}]{Loviat}%
  \BibitemOpen
  \bibfield  {author} {\bibinfo {author} {\bibfnamefont {F.}~\bibnamefont
  {Loviat}}, \bibinfo {author} {\bibfnamefont {H.~M.}\ \bibnamefont {Ronnow}},
  \bibinfo {author} {\bibfnamefont {C.}~\bibnamefont {Renner}}, \bibinfo
  {author} {\bibfnamefont {G.}~\bibnamefont {Aeppli}}, \bibinfo {author}
  {\bibfnamefont {T.}~\bibnamefont {Kimura}}, \ and\ \bibinfo {author}
  {\bibfnamefont {Y.}~\bibnamefont {Tokura}},\ }\href {\doibase
  10.1088/0957-4484/18/4/044020} {\bibfield  {journal} {\bibinfo  {journal}
  {Nanotechnology}\ }\textbf {\bibinfo {volume} {18}} (\bibinfo {year}
  {2007}),\ 10.1088/0957-4484/18/4/044020}\BibitemShut {NoStop}%
\end{thebibliography}%
\end{document}